\documentclass{aastex}          
\usepackage{spr-astr-addons}    
\usepackage{url}
\usepackage{cases}

\begin{document}
%
 \title{Solar cycle variations in the growth and decay of
sunspot groups} 

\shorttitle{Growth and decay of sunspot groups}
\shortauthors{J. Javaraiah}

\author{J. Javaraiah\altaffilmark{1}} 
     
\email{jj@iiap.res.in} 

\altaffiltext{1}{
   Indian Institute of Astrophysics, Bangalore- 560034, India}


\begin{abstract}
We analysed  the  combined  Greenwich
 (1874\,--\,1976)  and Solar Optical
Observatories Network (1977\,--\,2011) data on sunspot groups.
The daily rate of change of the area
of  a spot  group is computed using the differences between
the epochs of the spot group observation on
any two
consecutive days during its life-time and
between the   corrected whole
spot areas
 of the spot group at these epochs. Positive/negative value of
the daily rate of change of the area of a spot group
represents the growth/decay rate of the spot group.
We found that 
the total amounts of  growth
and  decay of spot groups whose life times $\ge 2$ days
 in a given time interval (say one-year)
 well correlate to the amount of activity in the same interval.
 We have also found that there exists a reasonably good correlation and
 an approximate linear 
relationship between  the  logarithmic values of the decay rate
and
   area of the spot group at the first day of 
the corresponding 
consecutive days, largely suggesting that a large/small area (magnetic flux) 
decreases in a faster/slower rate.  
There exists a long-term variation (about 90-year) in the 
slope of the  linear relationship. 
 The solar cycle variation in the decay of spot groups may have a 
 strong relationship with   the corresponding variations in solar 
energetic phenomena such as solar flare activity. The decay of spot groups 
 may also     
 substantially
 contribute  to  the
coherence relationship between the total solar irradiance and the solar
activity variations.
\end{abstract}


\section{Introduction}
Solar activity affects us in many ways.
Flares and coronal mass ejections pose a serious hazard to
 astronauts, satellites, polar air-traffic, electric power grids and
telecommunications facilities on short time-scales ranging from hours to  days.
The solar radiative output affects planetary and global climate on much longer
time-scales (from decades to stellar evolutionary time-scales).
The study of varitions in solar activity is important for understanding
the underlying mechanism of solar activity and for predicting the level of
activity in view of the  activity impact
on space weather and global climate~\citep{hat09}.

Recently,\citep[][hereafter Paper-1]{jj11a},
 using  the  combined  Greenwich (May/1874 to 1976) and Solar Optical
Observatories Network (1977\,--\,2009) data on sunspot groups, we studied
 the long-term variations in the daily mean percentage 
growth and decay  rates  of
sunspot groups.
Two of the main results found from this study are:
(i) From the beginning of Cycle~23
the growth rate is substantially decreased and   near the end
(2007\,--\,2008)
the growth  rate is lowest in the past about 100 years. (ii)
In the extended part (beyond the length of the declining part of a
normal cycle) of the
 declining phase of this cycle, the decay
rate steeply increased  and it is largest in the
 beginning of the current Cycle~24.
These unusual properties of the growth and the decay rates  during Cycle~23
  may be related to cause of the  very
  long declining phase of this cycle  with the
 unusually  deep and prolonged current minimum.
However,  no significant correlation was found either between  spot group
 growth rate
 and sunspot number or between the latter and  spot group decay rate.
The patterns of variations in the growth and decay rates are found to be
 considerably different
in different cycles. In fact, the mean percentage 
growth rate of the spot groups in the
declining phases of some cycles  is found to be considerably large
  and the mean percentage decay rate 
 of the spot groups in the rising phases of some cycles is found to be 
considerably large (see Fig.~6 in Paper-1).
In the present paper we showed that the annual amounts of growth and
decay of spot groups well correlated to
the annual amount of activity and discussed the implications 
of this result on the solar cycle variation of the solar energetic phenomena 
such such as 
solar flares and that of the total solar irradiance (TSI).
 We have also studied the dependence of   the growth and  decay rates 
of the spot groups on their sizes. Such a study may help for understanding the
underlying mechanism of the emergence/growth and decay of 
magnetic flux.

In the next section we  describe the methodology and the data analysis.
In Section~3 we  present  the results.
In Section~4 we  draw
conclusions and
briefly discuss the implications of them.

\section{Methodology  and data analysis}
 Here we have used the combined  Greenwich
(May/1874 to December/1976) and  Solar Optical Observation
Network (SOON) (January/1977 to May/2011) sunspot group
data, which are
 taken  from  David Hathaway's  website, 
 {\tt http://solarscience.msfc.nasa.gov/\break greenwich.shtml}.
These data
included the observation time (the Greenwich data contain the date with the
fraction of the day, in  the SOON data
 the fraction  is rounded to 0.5 day),
heliographic latitude  and
longitude, central meridian distance (CMD), and
corrected umbra and whole-spot areas (in millionth of solar hemisphere, msh), etc., of the spot
groups for each day of observation.
The
positions of the groups  are geometrical  positions of the
centers of the groups.
In case of SOON data, we increased area by a factor of 1.4.
 David Hathaway found this correction was necessary to
 have a combined homogeneous
 Greenwich and SOON
data (see aforementioned web-site of David Hathaway).
The area of a spot (or spot group) is closely connected with the
magnetic flux of the spot (or spot group).
 The 130 msh area
 $\approx\ 10^{22}$ Mx magnetic flux~\citep[$e.g.$, see][]{wan89}.

\begin{table*}[tp]
\label{tab1}
   \caption{Values ({\bf in msh day$^{-1}$}) of cycle $S_G$, $S_D$ and $S_A$ 
{\bf (in msh)} and the
corresponding number of data points k, m, and N, respectively.}

\scriptsize
   \begin{tabular}{lcccccccccccccc}
        \tableline
Year &$S_G$& k &$S_D$ & m &$S_A$& N &&Year &$S_G$& k &$S_D$ & m &$S_A$ &
 N\\
        \tableline
 1874$^{\mathrm a}$& 7704&  101& $-$10401&  185&  110848&   410&& 1943&  10559&  140&  $-$8958&  205&  100054&   450\\
 1875&   5796&   85&  $-$7008&  146&   71824&   376&& 1944&   3790&  106&  $-$4454&  130&   41732&   335\\
 1876&   4109&   57&  $-$5128&  121&   37802&   256&& 1945&  14531&  287& $-$15558&  450&  142271&  1008\\
 1877&   1925&   48&  $-$2591&   87&   30574&   218&& 1946&  52065&  816& $-$55259& 1193&  613676&  2617\\
 1878&    917&   18&  $-$1041&   37&    7753&    76&& 1947&  93043& 1293& $-$80703& 1808&  880814&  3961\\
 1879&   1447&   35&  $-$1664&   48&   12711&   114&& 1948&  64320& 1099& $-$71321& 1621&  660406&  3555\\
 1880&  16301&  225& $-$14819&  275&  148066&   713&& 1949&  73716& 1153& $-$71714& 1665&  709618&  3618\\
 1881&  23403&  384& $-$28231&  582&  233877&  1349&& 1950&  39326&  652& $-$40250&  999&  410458&  2133\\
 1882&  32410&  497& $-$37242&  666&  322272&  1518&& 1951&  31605&  487& $-$33303&  817&  383792&  1712\\
 1883&  35147&  509& $-$41690&  770&  392421&  1678&& 1952&  10865&  221& $-$15038&  439&  136097&   884\\
 1884&  35357&  620& $-$30821&  873&  348129&  1904&& 1953&   3698&   92&  $-$6301&  198&   48729&   399\\
 1885&  23591&  452& $-$26822&  661&  272557&  1438&& 1954&   2466&   40&  $-$1466&   56&   11528&   154\\
 1886&  10435&  228& $-$10716&  302&  124279&   692&& 1955&  15708&  292& $-$20003&  523&  188297&  1102\\
 1887&   4534&  104&  $-$6385&  217&   58856&   423&& 1956&  74858& 1033& $-$75226& 1625&  805335&  3495\\
 1888&   2034&   59&  $-$3676&  126&   30458&   254&& 1957&  92087& 1347& $-$97049& 2025& 1030795&  4519\\
 1889&   2420&   61&  $-$2002&   81&   26031&   181&& 1958& 104359& 1456&$-$107399& 2132& 1016484&  4649\\
 1890&   3680&   61&  $-$4633&  103&   33662&   245&& 1959&  85162& 1272& $-$88205& 1927&  966043&  4178\\
 1891&  20571&  344& $-$18502&  518&  182676&  1147&& 1960&  52810&  947& $-$56159& 1339&  543777&  2994\\
 1892&  35153&  611& $-$42382&  970&  400948&  2163&& 1961&  22120&  428& $-$22192&  630&  205291&  1474\\
 1893&  45905&  831& $-$51331& 1366&  489336&  2877&& 1962&  12848&  278& $-$16083&  436&  149997&   944\\
 1894&  40539&  723& $-$44901& 1197&  426001&  2544&& 1963&   8981&  209& $-$10188&  322&   94727&   749\\
 1895&  31791&  546& $-$30944&  907&  327530&  1992&& 1964&   3039&   98&  $-$2354&  139&   17462&   359\\
 1896&  19073&  337& $-$19692&  493&  179861&  1105&& 1965&   5326&  130&  $-$5271&  206&   38028&   473\\
 1897&  11089&  241& $-$12251&  369&  164355&   850&& 1966&  24612&  409& $-$21802&  553&  198533&  1377\\
 1898&  10413&  182& $-$12889&  353&  127138&   729&& 1967&  54784&  920& $-$51612& 1269&  511008&  3149\\
 1899&   3557&   90&  $-$5251&  192&   38288&   371&& 1968&  44987&  775& $-$52904& 1235&  525099&  2748\\
 1900&   3573&   76&  $-$3971&  122&   24558&   260&& 1969&  46869&  781& $-$46647& 1262&  489365&  2745\\
 1901&    962&   21&  $-$1150&   35&    9695&    72&& 1970&  51669&  941& $-$51765& 1359&  527659&  3112\\
 1902&   3014&   33&  $-$2297&   54&   20794&   128&& 1971&  27523&  572& $-$30248&  951&  329595&  2063\\
 1903&   7485&  163& $-$11062&  316&  111170&   703&& 1972&  25923&  609& $-$34400&  938&  302052&  2111\\
 1904&  18531&  373& $-$20863&  543&  163222&  1285&& 1973&  13521&  288& $-$16506&  496&  151689&  1111\\
 1905&  28320&  507& $-$35503&  778&  399858&  1812&& 1974&  16077&  298& $-$13854&  407&  130523&  1012\\
 1906&  26295&  497& $-$25688&  723&  255142&  1731&& 1975&   6249&  122&  $-$5548&  157&   56883&   440\\
 1907&  30918&  524& $-$30768&  781&  358190&  1810&& 1976&   5063&  125&  $-$5381&  162&   54738&   394\\
 1908&  22080&  458& $-$19575&  666&  228713&  1568&& 1977&  14471&  216& $-$15379&  240&  116909&   862\\
 1909&  22123&  413& $-$21320&  589&  231773&  1349&& 1978&  52217&  761& $-$54128& 1141&  471536&  3701\\
 1910&   9187&  210&  $-$9264&  324&   88160&   723&& 1979&  85123& 1172& $-$85067& 1649&  753027&  5025\\
 1911&   1964&   79&  $-$3276&  123&   21043&   263&& 1980&  77175&  945& $-$72511& 1219&  736414&  3672\\
 1912&   2203&   50&  $-$1497&   54&   12676&   146&& 1981&  81318&  972& $-$77640& 1243&  777174&  3645\\
 1913&    347&   15&   $-$466&   22&    2683&    60&& 1982&  78540&  815& $-$79842& 1119&  756462&  3072\\
 1914&   4810&  107&  $-$5555&  165&   51100&   408&& 1983&  36456&  482& $-$35434&  662&  318990&  2069\\
 1915&  21391&  456& $-$23446&  685&  235067&  1738&& 1984&  25998&  293& $-$25746&  414&  274008&  1263\\
 1916&  27793&  585& $-$27313&  881&  237122&  2138&& 1985&   7350&   95&  $-$7868&  161&   61432&   502\\
 1917&  52994&  982& $-$51555& 1437&  515459&  3266&& 1986&   4732&   76&  $-$5264&  104&   41160&   347\\
 1918&  38390&  766& $-$39210& 1177&  365995&  2689&& 1987&   9716&  171& $-$10808&  244&  100296&   793\\
 1919&  30157&  588& $-$31948&  941&  349358&  2082&& 1988&  43022&  560& $-$45724&  778&  464660&  2453\\
 1920&  18914&  343& $-$19174&  578&  202831&  1289&& 1989&  86744&  921& $-$90216& 1308&  885234&  3935\\
 1921&  13040&  234& $-$12999&  376&  137361&   869&& 1990&  71932&  948& $-$72744& 1304&  700798&  3926\\
 1922&   8947&  141&  $-$8697&  227&   81593&   501&& 1991&  77476&  898& $-$80710& 1311&  843920&  3840\\
 1923&   2381&   54&  $-$2900&  104&   19981&   239&& 1992&  50960&  630& $-$46186&  858&  460789&  2470\\
 1924&   8167&  142&  $-$9089&  248&   90535&   543&& 1993&  23436&  374& $-$21490&  442&  235760&  1274\\
 1925&  29665&  460& $-$25162&  692&  280126&  1600&& 1994&  11942&  214& $-$11228&  281&  115262&   836\\
 1926&  31973&  584& $-$36563&  914&  420728&  2062&& 1995&   7420&  130&  $-$7966&  187&   54502&   492\\
 1927&  37578&  621& $-$35608& 1007&  355199&  2194&& 1996&   4564&   55&  $-$3108&   64&   27972&   217\\
 1928&  38609&  661& $-$44095& 1114&  459454&  2399&& 1997&   9702&  140&  $-$7602&  160&   71022&   534\\
 1929&  36813&  639& $-$39030& 1014&  414790&  2244&& 1998&  30044&  419& $-$23450&  483&  258972&  1568\\
 1930&  16662&  367& $-$20441&  590&  172294&  1295&& 1999&  46578&  629& $-$37016&  727&  393680&  2236\\
 1931&  11866&  204& $-$10740&  303&   89798&   712&& 2000&  60830&  870& $-$53970& 1029&  544936&  2991\\
 1932&   5189&  120&  $-$4309&  177&   52896&   434&& 2001&  57722&  856& $-$53794& 1081&  578474&  3040\\
 1933&   1930&   63&  $-$3292&   93&   29935&   211&& 2002&  66318&  898& $-$59583& 1110&  620455&  3134\\
 1934&   3158&   81&  $-$4404&  137&   38006&   308&& 2003&  41055&  544& $-$36260&  719&  372722&  1910\\
 1935&  22155&  359& $-$19788&  531&  208639&  1189&& 2004&  25410&  344& $-$23849&  427&  230916&  1153\\
 1936&  41495&  719& $-$42964& 1079&  386772&  2500&& 2005&  20244&  241& $-$18200&  332&  180222&   864\\
 1937&  68489& 1024& $-$63853& 1538&  700115&  3491&& 2006&   8680&  135&  $-$7615&  186&   84127&   538\\
 1938&  55398&  872& $-$57529& 1440&  676297&  3154&& 2007&   3430&   60&  $-$5124&   97&   46088&   282\\
 1939&  50506&  821& $-$53038& 1233&  535160&  2693&& 2008&   1246&   26&  $-$1456&   42&    7805&   111\\
 1940&  38560&  612& $-$35387&  873&  348600&  1947&& 2009&   2044&   33&   $-$910&   24&    9058&   115\\
 1941&  23683&  396& $-$23130&  620&  223784&  1384&& 2010&   8644&  148&  $-$8428&  169&   72804&   528\\
 1942&  16901&  295& $-$16763&  426&  144345&   939&& 2011$^{\mathrm a}$&  10129&  136&  $-$7868&  126&   73178&   471\\
        \tableline
\end{tabular}
 \tablenotetext{a} {indicates  data are available only  for about half
 years in the cases of 1874 and 2011.}
\end{table*}

\begin{figure*}[t]
\includegraphics[width=\textwidth]{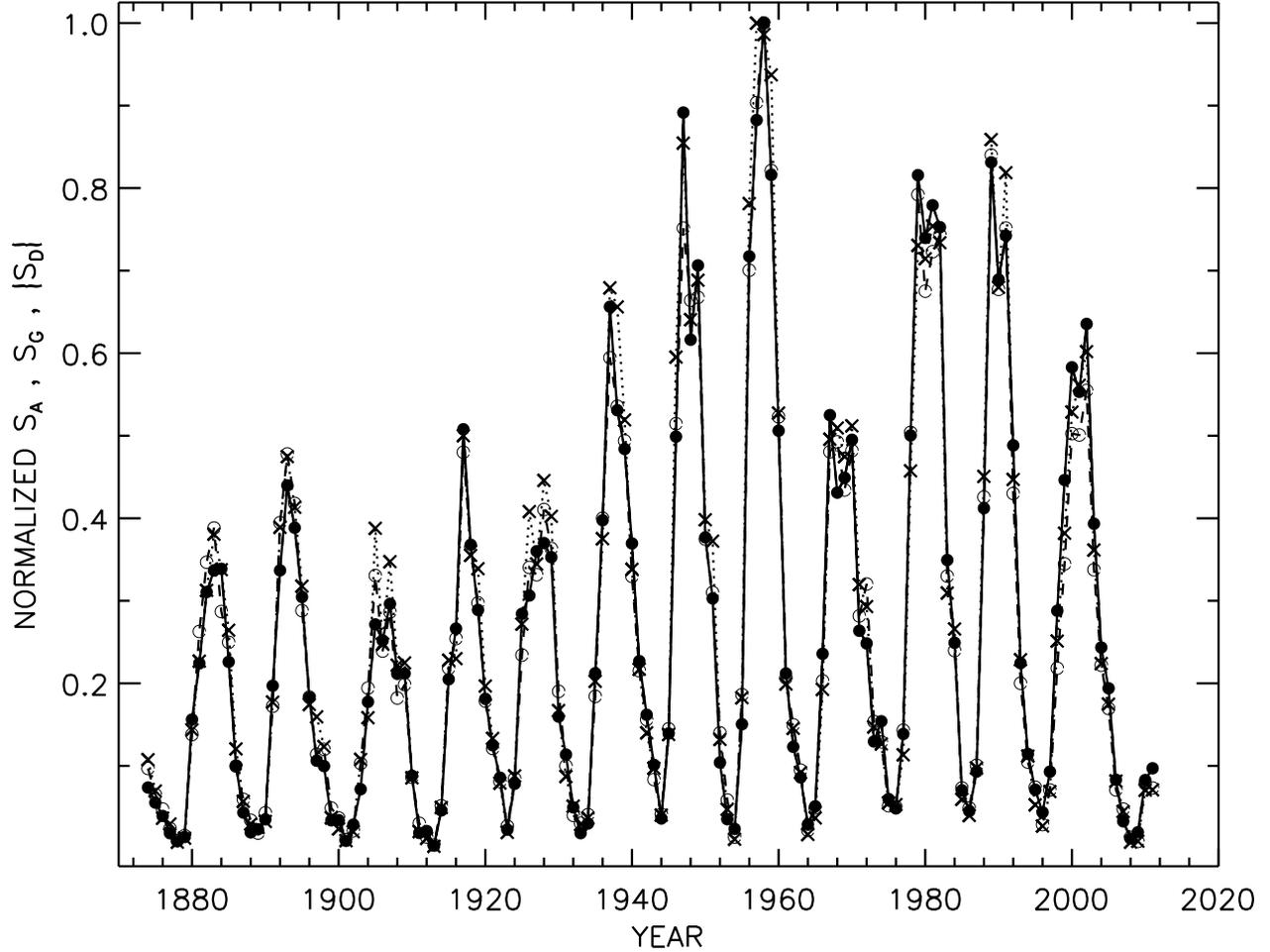}
\caption{Plots of normalized
 annual sum of daily area ($S_A$,
 cross-dotted curve),
  annual amounts of growth ($S_G$, filled circle-solid curve)
and decay ($S_D$,
open circle-dashed  
curve) of spot groups {\it versus} year. The $S_A$ , $S_G$ and $S_D$ are
normalized with their respective maximum values (cf., Table~1)}
\label{fig1}
\end{figure*}

\begin{figure}[t]
\includegraphics[width=\columnwidth]{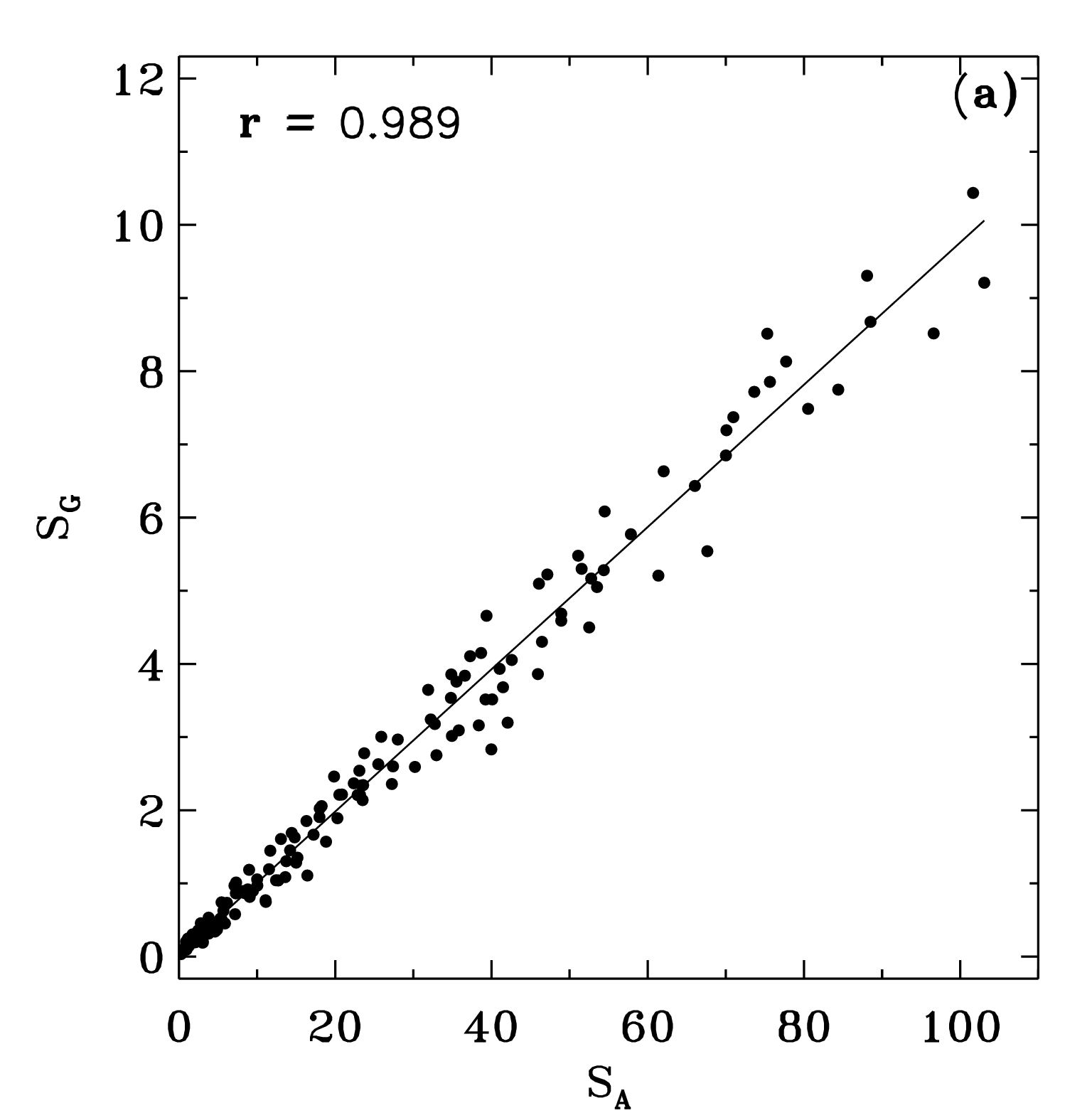}
\includegraphics[width=\columnwidth]{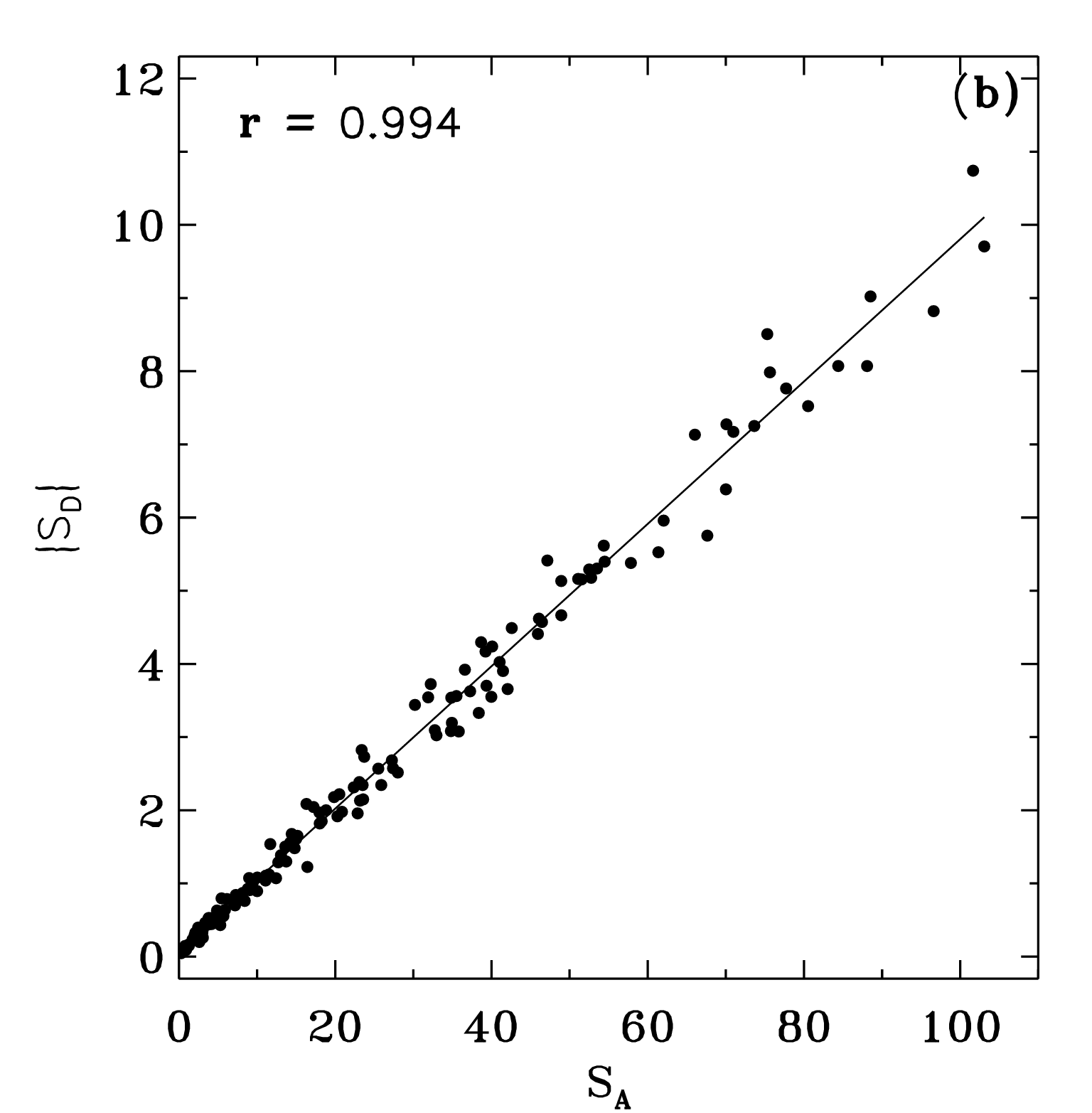}
\caption{Scatter plots of the annual
  $S_A$ and $|S_D|$ {\it versus}
 $S_A$ (the values which are given in Table~1 and divided by $10^4$).
The Solid line represents the corresponding linear
 relationship and the value of the corresponding  correlation coefficient (r)
is also shown}
\label{fig2}
\end{figure}

\begin{figure}[t]
\includegraphics[width=\columnwidth]{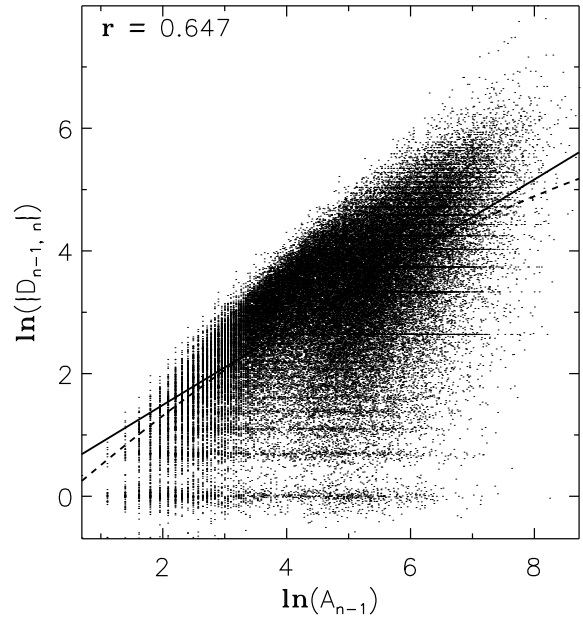}
\caption{Plot of $\ln(|D_{\rm n-1, n}|))$
{\it versus}  $\ln(A_{\rm n-1})$.
 $D_{\rm n-1, n}$ and $A_{\rm n-1}$ ($cf.$, \ref{eqn1} and \ref{eqn2}) are  determined
 from the data during the whole
period 1874\,--\,2011
(it should be noted that the values of n is different for
the growth and decay rates).
The continuous and dashed curves represent the corresponding linear
and quadratic  fits, receptively}
\label{fig3}
\end{figure}

\begin{figure}[t]
\includegraphics[width=\columnwidth]{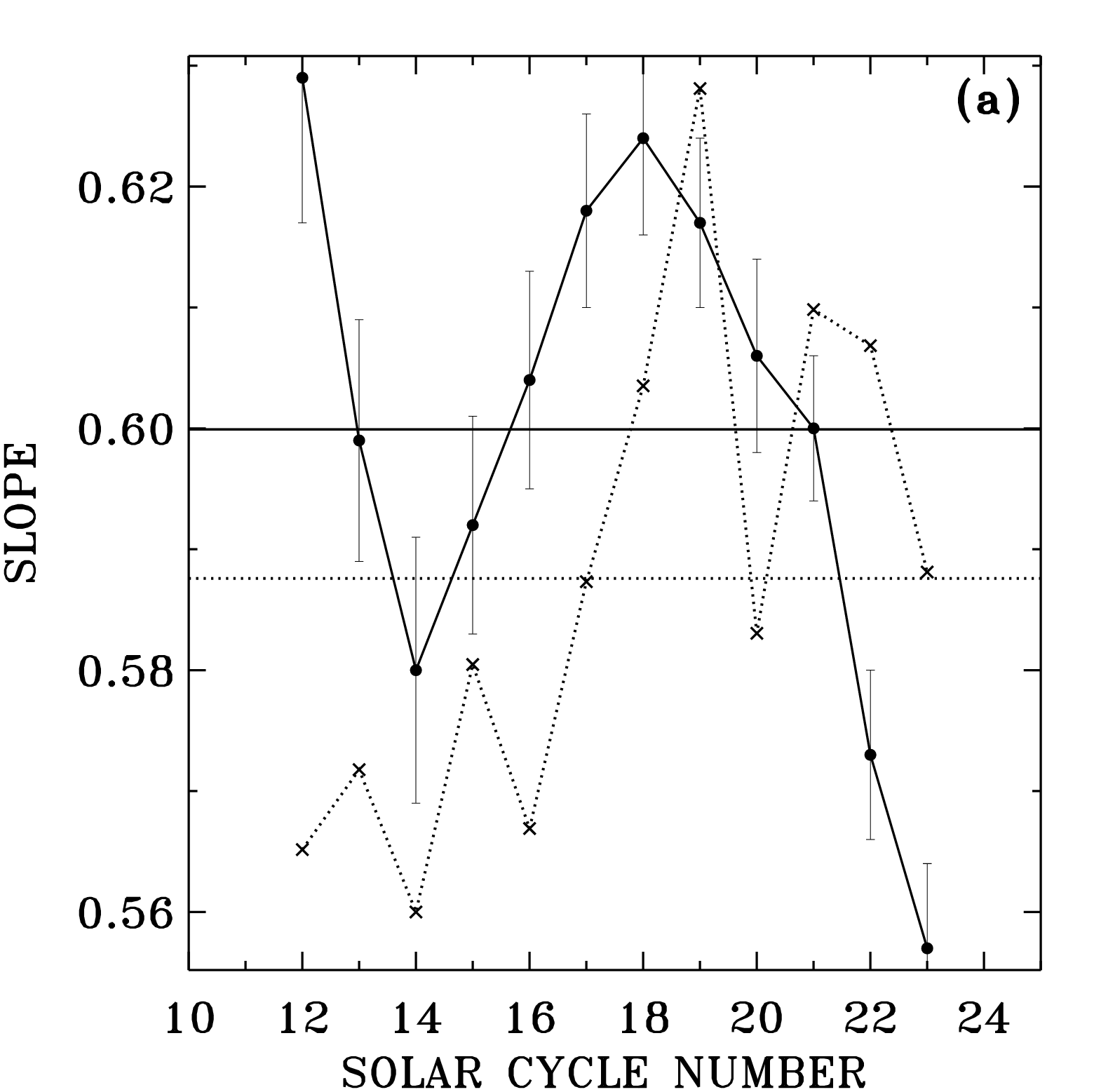}
\includegraphics[width=\columnwidth]{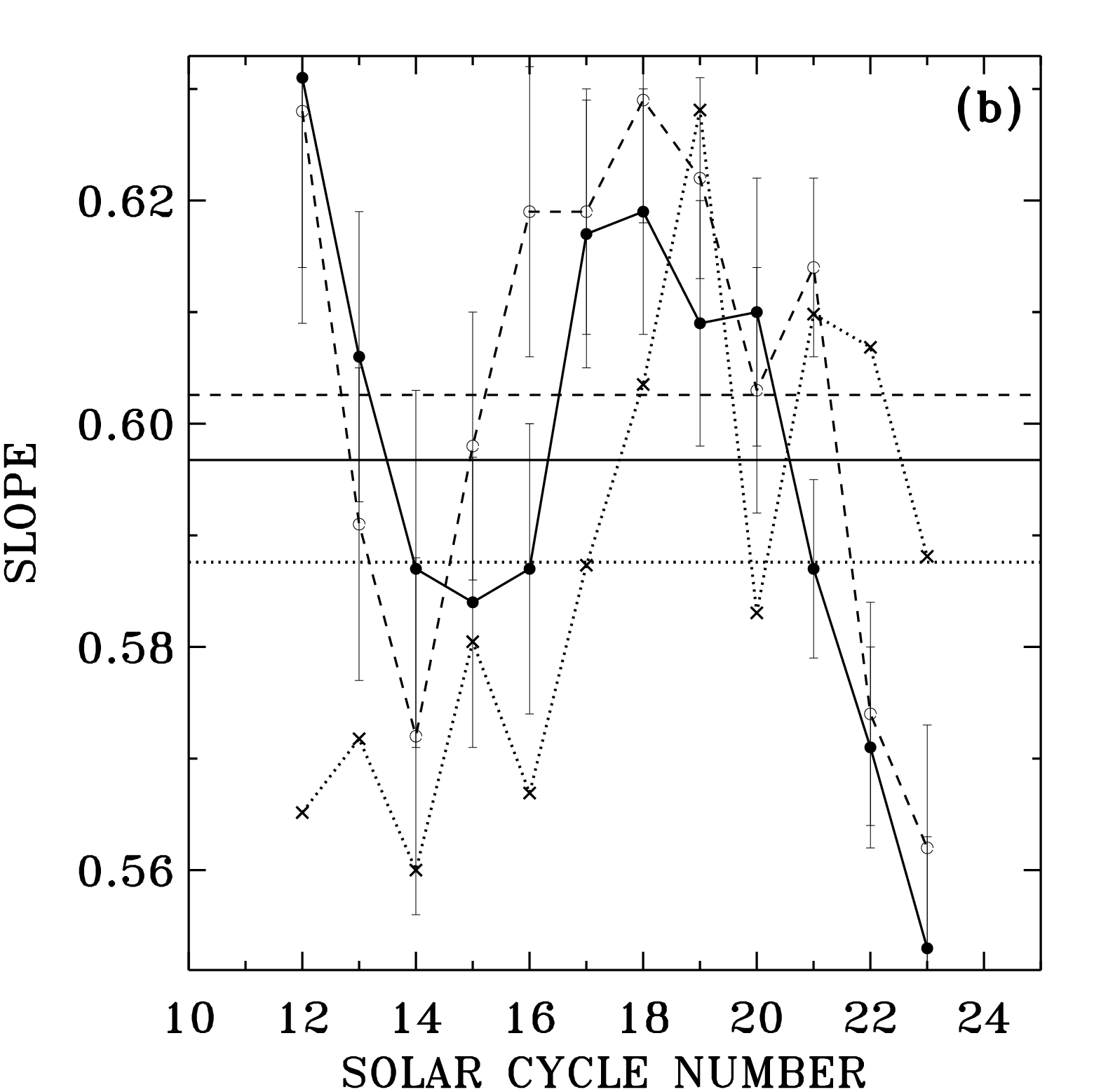}
\caption{Plots of cycle-to-cycle variation in the slope  
 of the linear relationships between $\ln(|D_{\rm n-1, n}|))$
{\it versus}  $\ln(A_{\rm n-1})$ derived from the data of 
spot groups in the whole sphere (upper panel) and in  the 
different hemispheres
(lower panel):  
northern hemisphere (open circle-dashed curve) and 
southern hemisphere (closed circle-solid curve). 
The cross-dotted curve represents the variation in the amplitudes of
the sunspot cycles 12\,--\,23 (normalized to the scale of the slope). 
The corresponding  values of the mean (over all cycles) 
 are indicated by the horizontal lines of the receptive type}
\label{fig4}
\end{figure}

\begin{figure}[t]
\includegraphics[width=\columnwidth]{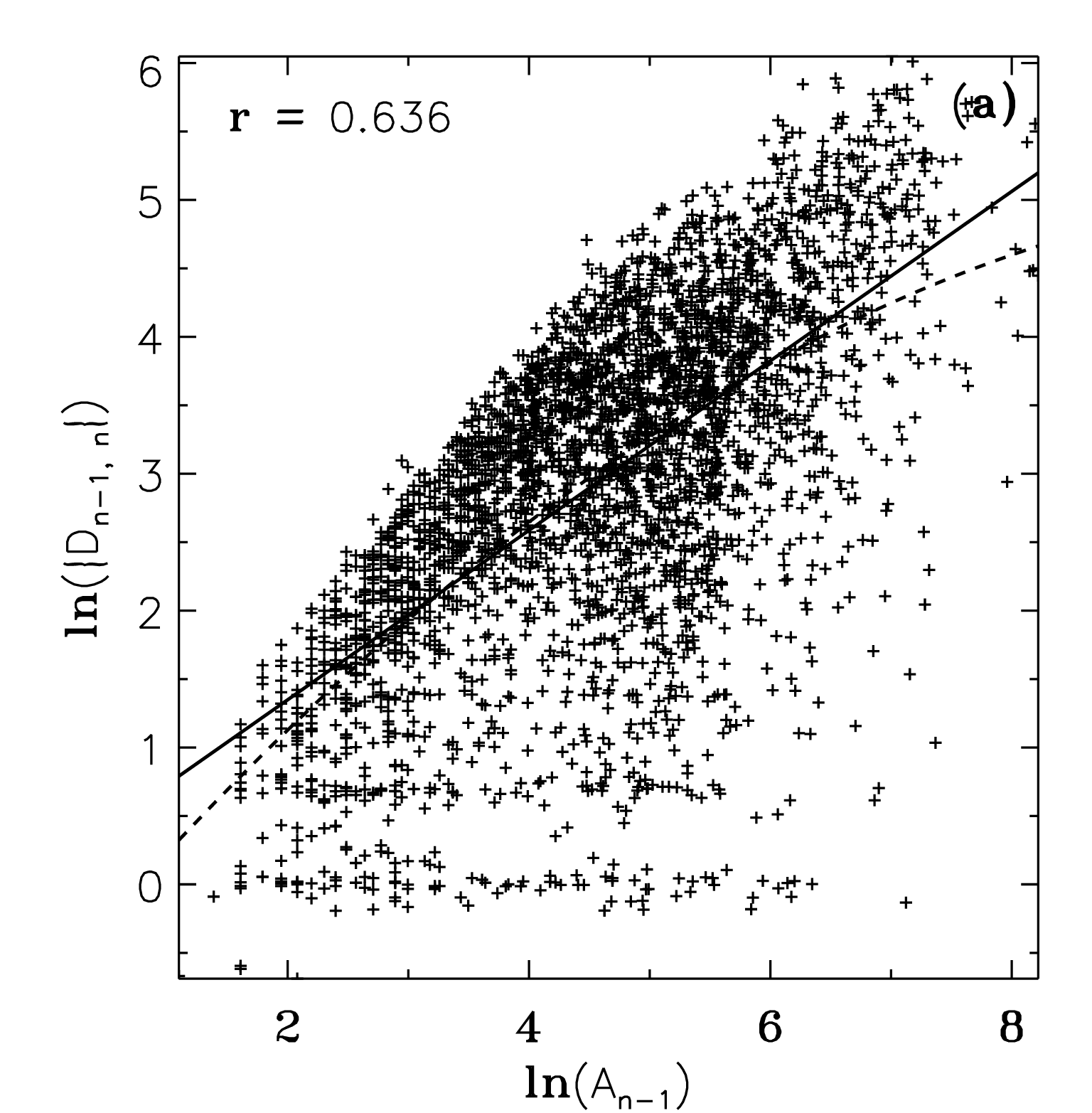}
\includegraphics[width=\columnwidth]{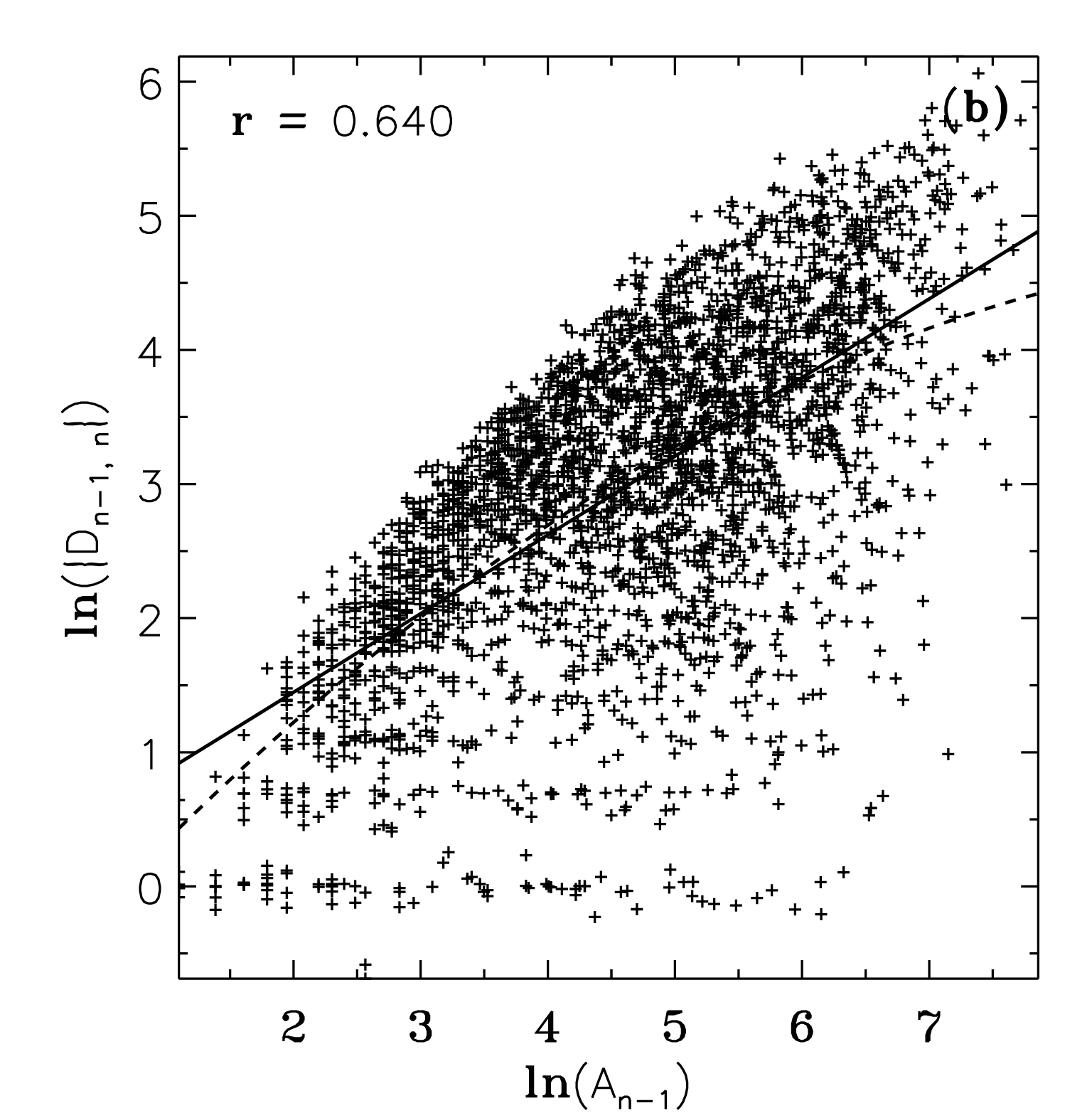}
\caption{The same as Figure~3, but determined separately from the data of 
the spot groups in the northern 
(upper panel) and
 southern (lower panel) hemispheres  during Cycle~16 only} 
\label{fig5}
\end{figure}

\begin{figure*}[t]
\includegraphics[width=\textwidth]{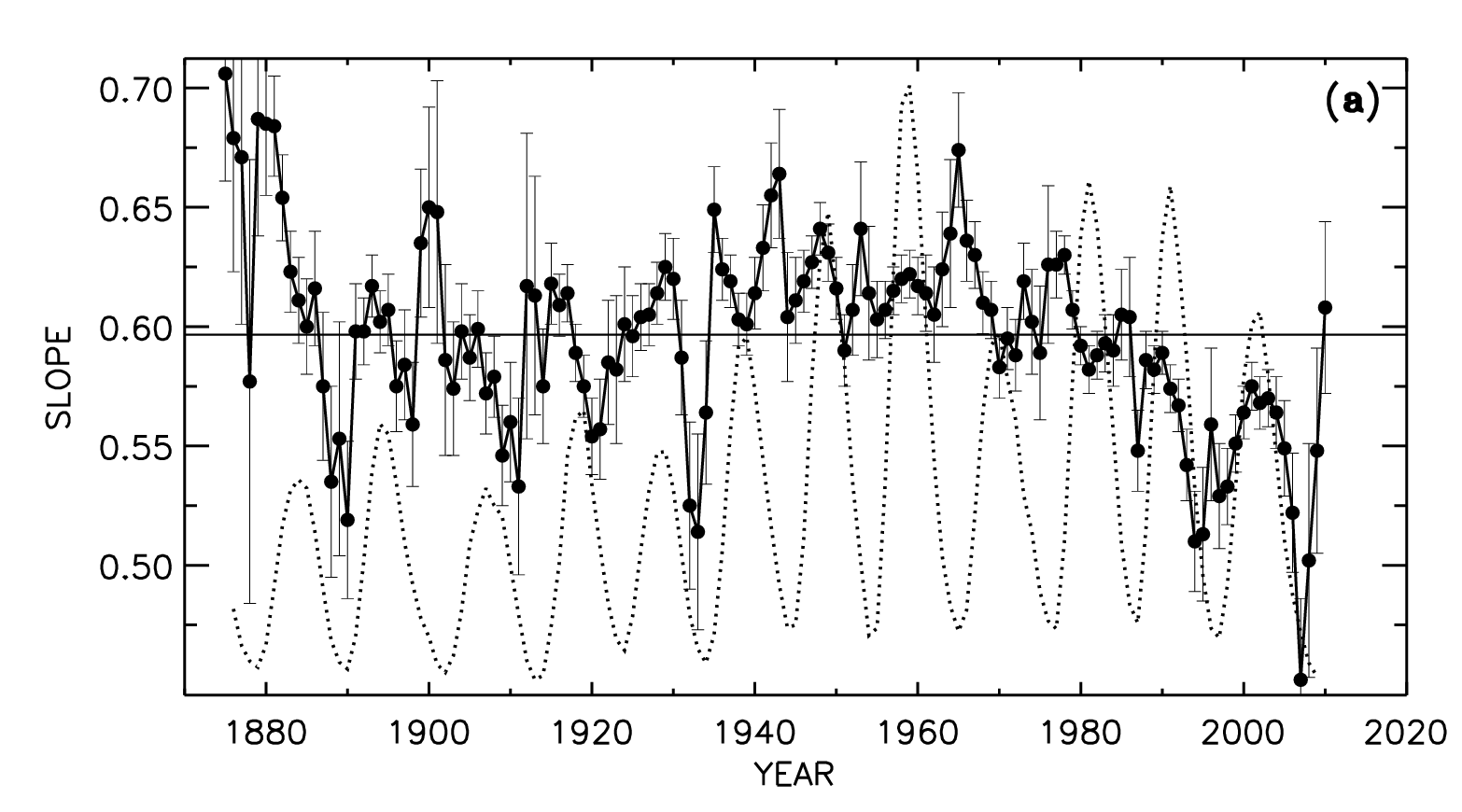}
\includegraphics[width=\textwidth]{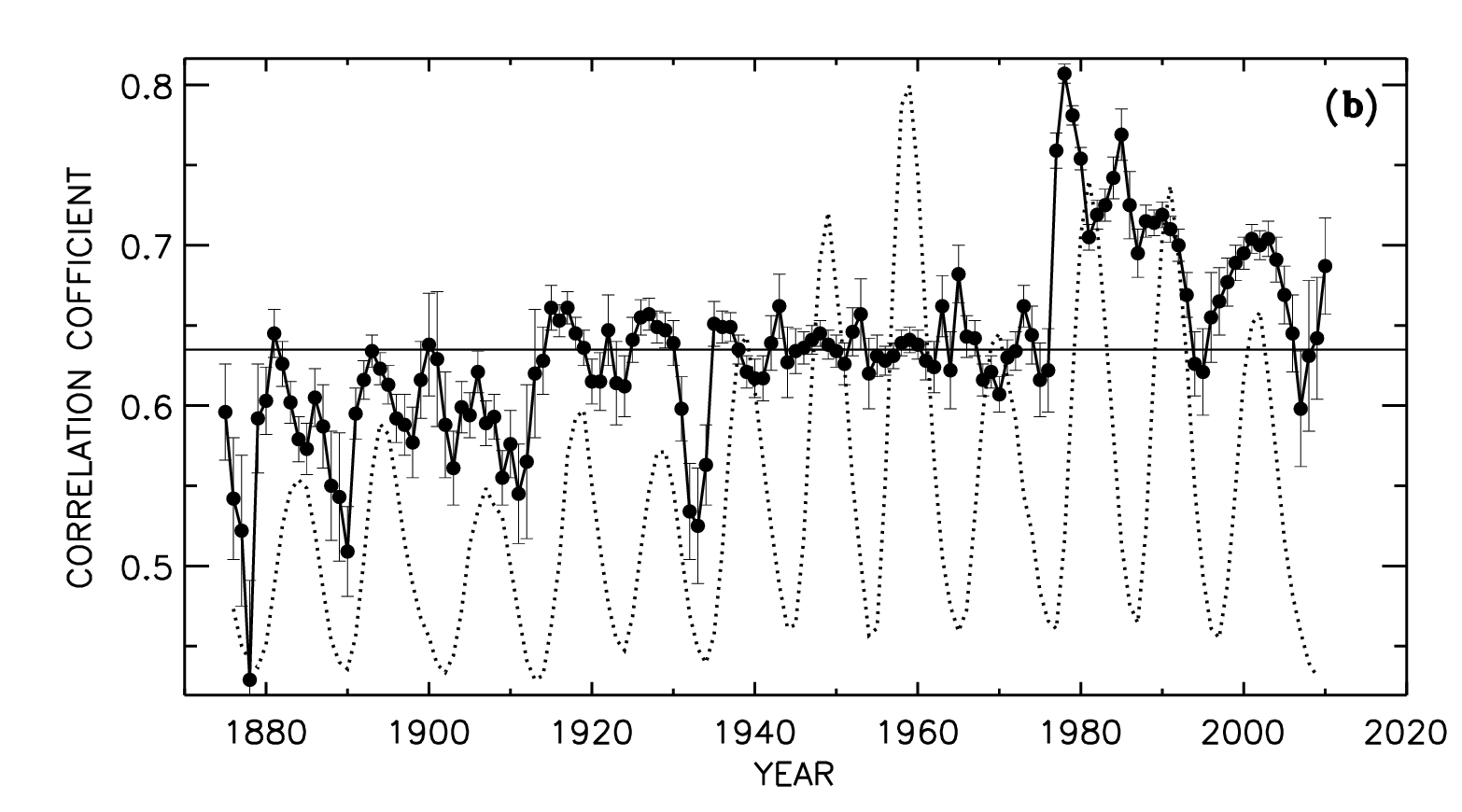}
\caption{Plots of the varitions in the slope (upper panel) 
 of the linear relationship between $\ln(|D_{\rm n-1, n}|))$
and $\ln(A_{\rm n-1})$, and 
the corresponding correlation coefficient (lower panel),
 derived from the whole sphere spot group data in 3-year MTIs 
1874\,--\,1876, 1875\,--\,1877,....,2009\,--\,2011,  
{\it versus} the middle year of the interval.  
The dotted curve represents the variation in the yearly 
 international sunspot number during 1874\,--\,2009, smoothed by taking  
 3-year running average (normalized to the scales of the slope and 
the correlation coefficient) .
 The mean values of the slope and the correlation coefficient, 
over the whole duration,    are indicated by 
the horizontal solid lines}
\label{fig6}
\end{figure*}

\begin{figure}[t]
\includegraphics[width=\columnwidth]{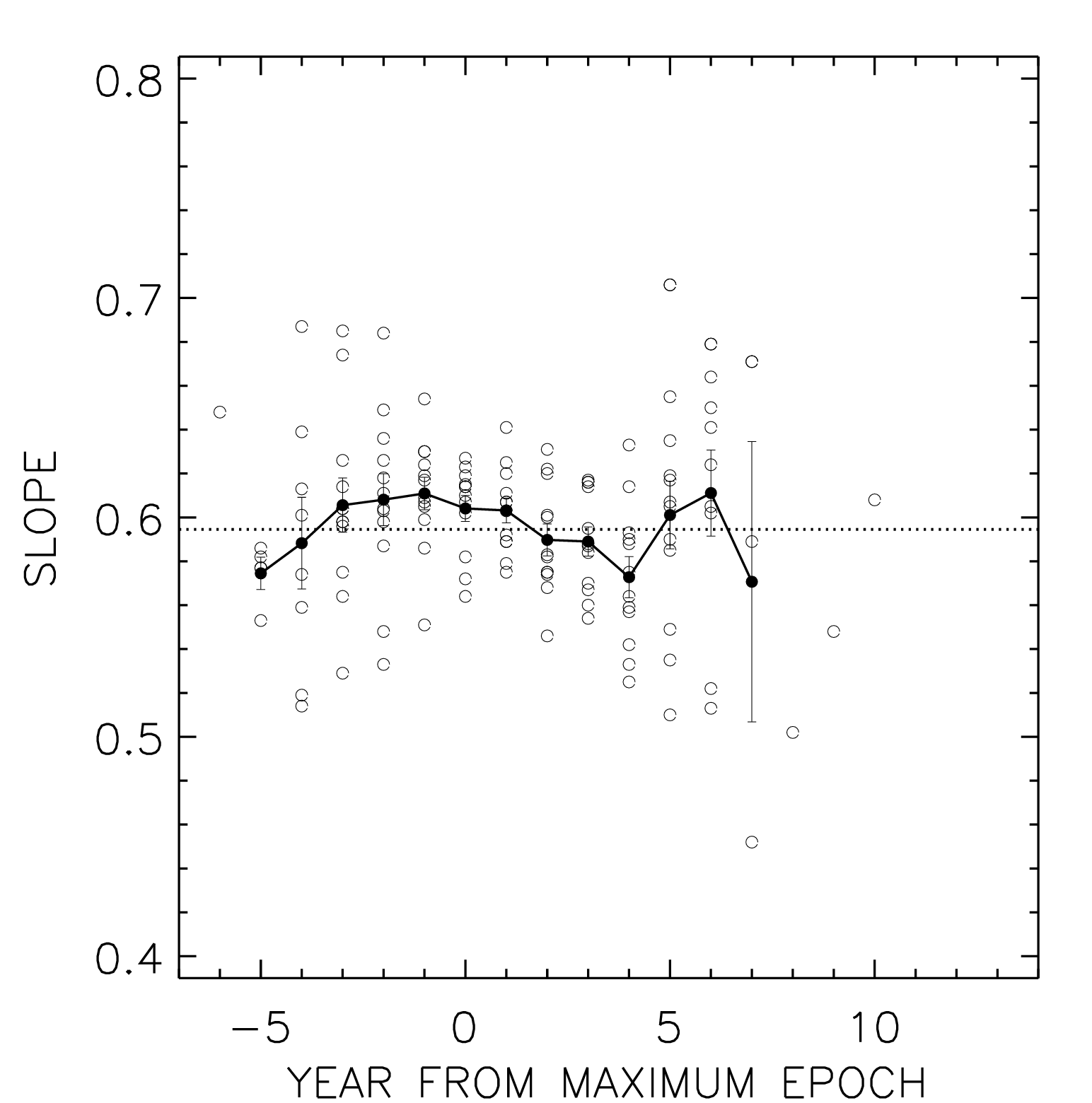}
\caption{Plots of the  mean slope values  in 3-MTIs, i.e., 
the data shown in Fig.~\ref{fig6}(a), versus  the year of the 
solar cycles, 11\,--\,24. The filled-circle-continuous curve represents 
the mean solar cycle variation determined from the mean values in 
3-MTIs. The error bar represents the standard error. There are only one data
 point at years -6 (begin of Cycle~14), 8 (end of Cycle~23)  and 9\,--\,10
 (begin of Cycle~24)} 
\label{fig7}
\end{figure}

The data reduction and analysis are similar as described  in Paper-1.
A brief  descriptions of them is given here (for details see Paper-1).
We have taken all the precautions which were taken in Paper-1.
We have used the corrected daily whole-spot areas
(umbral value + penumbral value) of spot groups ($A$).
A spot group is included when the  observations of it are available for two
or more consecutive days. The spot groups having the group numbers with 
suffix A, B, etc., which are available in the SOON data for some years, 
are not used in this analysis. However, they are few and  many of 
them found to have zeros  for values  $A$.  
The daily rate of change of the area ($\frac{\Delta{A}}{\Delta{t}}$) of
  a spot  group is computed using the differences between
the epochs of its observation
in
consecutive days and between the   corrected whole spot areas
 of the spot group at these epochs. That is,
\begin{equation}
\label{eqn1}
\frac{\Delta{A}}{\Delta{t}} = 
\frac{A_{\rm n}-A_{\rm n-1}}{t_{\rm n}-t_{\rm n-1}},
\end{equation}

\noindent where $t$ is the epoch of observation  during
the life-time ($T$) of the spot group and ${\rm n} =$ 2, 3,...., $T$.
 The data correspond to only 0.5$> \Delta{t}\ < 2$ day are used. 
Positive and negative values of $\frac{\Delta{A}}{\Delta{t}}$  correspond to
 the
daily rates of  growth ($G$)  and decay ($D$) of the spot group,
 respectively, i.e., 

\begin{subnumcases} {\label{eqn2}  \frac{\Delta{A}}{\Delta{t}} =}
 > \ 0  &  $=\  G$, \\
  < \ 0 &  $=\  D$.
\end{subnumcases}

The   sums
 $S_G$ and $S_D$ of $G$ and $D$, $i.e.$ the total amounts of growth and decay
of the spot groups   in a given time interval,  are determined as follows:
\begin{equation}
\label{eqn3}
 S_G = \sum G_i\ \ {\rm and}\ \  
S_D = \sum D_j\ , 
\end{equation}
\noindent where $i=$ 1, 2,....,k  and $j=$ 1, 2,....,m;
 k and m   are
the number of data points of $G$ and $D$, respectively,
in the interval. We also determined the sum of the areas of spot
groups ($S_A$) in corresponding time interval, 
$S_A = \sum A_l$, $i.e.$,
the annual sum of $A$, where $l=$ 1,2,...,N.  N includes all reliable
data points (including those correspond to the spot groups which had  born
and dead within one day, which are available in Greenwich data).

We  determined  the annual
$S_G$ and $S_D$ and their correlations with annual $S_A$.
Obviously more contributions to
$S_G$ and $S_D$  are
coming from  the spot groups
 before and after reaching their
maximum areas, respectively (see Fig.~1 in Paper-1).
 It should be noted here $S_A$ and $S_D$ 
 not only did not include the corresponding contributions 
of other  activity phenomena, they even did not include the 
corresponding contributions of one-day groups. Therefore, 
they do not represent  the amounts of 
 complete  growth and decay  of magnetic activity in  a given
 time interval.

 To check whether the growth and decay rates of the spot groups 
 depend on the sizes of the spot groups 
we determined  the correlations and the linear/quadratic
list-square fit to logarithmic values of   
$G_{\rm n-1, n}$ and  $|D_{\rm n-1, n}|$, which are 
 derived from $A_{\rm n-1}$ and $A_{\rm n}$,  and corresponding  
 $A_{\rm n-1}$ in a given time interval
(it should be noted that the values of n are different for
the growth and decay rates).
The statistical significances of the 
 regression relations and the values of correlation coefficients are tested 
using $\chi^2$ and Student's 't' distributions, F-test, standard error and 
z-transformation tests~\citep{yk58}.

\section{Results}
\subsection{$S_G - S_A$ and $S_D - S_A$ relations}
In Table~1 we have given the annual values of $S_G$, $S_D$ and $S_A$.
Fig.~\ref{fig1} shows the plots of the normalized annual values of  $S_A$,
 $S_G$ and  $|S_D|$ 
 {\it versus} year.  It may be worth to note here that in cycle~23  $S_A$ 
has maximum at year 2002. The maximum of the international sunspot number 
(not shown here) is at 2000. Recently, \citet{ramesh10} found that the  
maximum of coronal mass ejection (CME) is close to  that of  the  sunspot 
area. 
Fig.~\ref{fig2} shows the correlations between
$S_G$ and $S_A$ and between $|S_D|$  and $S_A$
(all these  are divided by $10^4$).
These correlations,  correlation coefficients
$r = 0.989$ and $r =0.994$,   are
 very high (significant on $> 99.9$ confidence level).
The slopes of the  corresponding linear relationships
are almost equal ($0.097 \pm 0.001$)~\footnote{The $S_A - S_G$ and $S_A - S_D$  relations
 can be written as follows:\\ 
\begin{equation}
\label{eqn4}
{\bf
 \frac{S_A}{S_G} \approx \frac{S_A}{S_D}  \approx 10 \  {\rm day} .
}
\end{equation}\\
 It is well believed that large spot groups also  live long. 
In fact, there is a rule of the  proportionality of
the maximum area ($A^0$) of a sunspot
group to its life-time (T) (first plotted by Gnevyshev, 1938; and 
formulated by Waldmeier, 1955; see also Petrovay \& Van Driel-Getztelyi, 1997):
\begin{equation}
\label{eqn5}
 \frac {A^0}T \approx   10\ {\rm msh\ day}^{-1}.
\end{equation}\\
Incidentally, the right hand sides of (\ref{eqn4}) and (\ref{eqn5})
  have a value approximately equal to ten, but their units differ (the unit of former is day, whereas
 the unit of the latter is msh day$^{-1}$). They seem to be  independent relations.  
  The (\ref{eqn5}) 
represents the
average property of a whole individual spot group (it may not 
applicable to the  growth and the 
decay  portions of the spot groups, independently), 
whereas the (\ref{eqn4})
represents largely a  global property of the solar cycles.
}
 and  are statistically
highly significant
({\bf the values of Student's `t' are found to high, i.e. the values of their 
corresponding probabilities are found be close to one}).
 These relations  suggest that a larger
  amount of growth or
 decay  is associated with a larger
 amount of activity.
That is, the amounts of growth and decay of the flux in a given
time interval  depends on (proportional to) the total amount of
flux in that interval.
 Obviously the amounts of growth and decay
of magnetic flux is larger  during the
 maximum epochs than during minimum epochs.
During Maunder minimum $S_A$ was very low/absent. Hence, $S_G$ and $S_D$
were also very low/absent.

\subsection{Decay law}
\cite{how92} analysed Mt. Wilson sunspot and sunspot group data 
 during 
the period 1917\,--\,1985 and  found that the daily percentage 
umbral areas increases for small groups are larger in 
absolute terms than the percentage umbral areas decreases, 
and for large groups the daily 
percentage decreases are larger than increases.  
Fig.~\ref{fig3} shows the dependence of the   decay 
rate, $|D_{\rm n-1, n}|$, of a spot group determined 
from  $A_{\rm n-1}$ and $A_{\rm n}$ i.e.,
the areas at (n$-$1)th and nth days
during the life time of the spot group, on $A_{\rm n-1}$.
In this figure the continuous and dashed curves represent the corresponding
 linear and 
quadratic  fits.
The values of the intercept and the slope of the linear fit~\footnote{It is
 a power-law, $|D_{\rm n-1, n}| \approx e^{0.26} A_{\rm n-1}^{0.613}$, 
suggesting that 
 decay of spot group do not completely depend linearly on the size of
 the group.
It should be noted that the linear decay law is widely accepted 
as only a simplest 
approximation~\citep[see][]{pet97}.} 
 are $0.26\pm 0.02$ and $0.613 \pm 0.002$, respectively.
 Since  the number of data points, $ m =  88490$,     
over the whole period
from May/1874 to May/2011 
 is are very high,
 the levels of the
statistical significances of the values of 
the coefficients of both the linear and the quadratic fits (whose values are
not given here)  are very high.
 The $\chi^2$ value of both the linear and the quadratic fits are found to be 
insignificant at 5\% level. The $\chi^2$ value (68386) of the quadratic fit  
  is found to be slightly less than that (68649) of the linear fit. However, 
 the F-test indicated that no significant difference in the variances 
of these fits.
The correlation  between $|D_{\rm n-1, n}|$ 
   and
$A_{\rm n-1}$
 is  high ($r=0.65$)
 and statistically very significant (i.e., it is about 
19 times larger than  its standard error ($\approx \frac{1}{\sqrt{m}}$), 
and it is also found to significant 
in z-transform test).
  This and the linear (or  quadratic) relations  indicate that
a large/small flux  decays in a faster/slower rate. 
This is  consistent with the  $S_A - S_D$ relation above.

 In order to check the consistency of the aforementioned  analysis for
small data samples
and  dependence (if any) of the above found relationship 
on the solar cycle, 
  we  analysed the data of  individual cycles and 
also yearly data as well as 
 binning the  data into 
3-year moving time intervals (3-MTIs) successively shifted by one-year. 
In Fig.~\ref{fig4}, we show the cycle-to-cycle variation in the slope 
 of the linear relation between $\ln (|D_{\rm n-1, n}|)$ 
and $\ln (A_{\rm n-1})$   derived  from  the individual cycles' 
 the whole sphere (disk) data,    and also separately from 
 the  northern and southern hemispheres' data,   of spot groups 
during each of the cycles, 12\,--\,23. 
 In the same figure we have also shown 
the variation in the amplitudes of the same cycles 
(the maximum monthly mean  
international sunspot numbers which are taken from the website, 
{\tt ftp//ftp.ngdc.noaa.gov/STP/Solar\_DATA/SUNSPOT\_\break NUMBERS}).
In each cycle  the values of the coefficients of  both the 
linear and quadratic laws 
as well as  the value of the  correlation coefficient  are
 found to be statistically significant and  the F-tests suggest that
 no significant differences in the variances  of the linear 
and quadratic fits. The ranges of the  correlation coefficient 
(and of the ranges
 of the 
corresponding number of data points, i.e.,  m values) are 
 0.592\,--\,0.608 (4547\,--\,11113), 0.582\,--\,0.596 (2033\,--\,6388), and  
0.599\,--\,0.609 (2514\,--\,4725)  
 determined from the whole sphere,   northern hemisphere and southern 
hemispheres' data, respectively,   during the cycles  12\,--\,23.
 In Fig.~\ref{fig4}  it can be seen that the slope 
varies  on a long-time scale of about 90-year, with amplitude of about 
0.02 unit.
{\bf The variation in each  hemisphere is very closely resemble to that of 
the whole sphere. There is a suggestion that in a large number of cycles 
the slope is steeper in the northern hemisphere than in the
 southern hemisphere. However, only  in  cycles~16 and 21 
 the differences between the values of the northern and 
  southern hemispheres are somewhat large and statistically significant.} 
 Fig.~\ref{fig5} shows the relationship between 
 $|D_{\rm n-1, n}|$ and 
  $A_{\rm n-1}$ 
in Cycle~16. The patterns of these relations are closely resemble to 
that   derived from for the whole 138 year data (cf., 
 Fig.~\ref{fig3}). The aforementioned  difference in the slopes between 
northern and southern hemisphere during Cycles~ 16 and 21, i.e., in a gap of
 about 5 cycles,  may be related to a  
44\,--\,55 year cycle in solar activity~\citep{yk93,jj08}.

The value (it is only 0.08) of the coefficient of the
 correlation between the sunspot activity
  (amplitude of the cycle) and the slope is found to be negligible. 
During the 90-year cycle  (Gleissberg cycle) in the sunspot activity
there are many relatively small time-scale strong fluctuations, whereas 
there are no such fluctuations during the 90-year cycle in the slope.
However, there is a close agreement in the epochs of the maxima 
and the minima of these  cycles of the slope and the cycle amplitude
(the phase shift between these is not clear in Fig.~\ref{fig4}). This may  
suggest the existence of a relationship between the long-term 
variations in  the slope and the sunspot activity.  

 Fig.~\ref{fig6} shows the   
variations in the slope and the  correlation coefficient determined from 
the data in 3-year MTIs 1874\,--\,1876, 1875\,--\,1877, ...., 2009\,--\,2011.
In order to check  the solar cycles trends  in these parameters, in this figure we have 
also shown the 
variation in the international sunspot number smoothed by taking 3-year 
running average.  
Only a few of these values of the slopes shown in this figure are  
 statistically insignificant. That is,  in many intervals the values of 
 $\chi^2$  are found to be insignificant at 5\% level, 
 and the Student's `t' tests suggest that the significant  levels of the 
corresponding  values of the correlation coefficient  are also good 
(Note:  the big jump of the correlation coefficient  from the interval 
1977 to 1978 (the high values from 1978 onward) could be just an artifact
 of the multiplication of the 
area values of the SOON data with 1.4, in order to  
  have a continuous 
and homogeneous data for the whole period
1874\,--\,2011 (cf., Sec.~2).
 As can be seen Fig.~\ref{fig6}(a) the values of the slopes  
 are considerably low near the declining ends  of small cycles 
(12, 16 and 23), particularly
in the end of Cycle 23  the slope is smallest in the last about 100 years
(which may be related to the unusually low and prolonged recent activity minimum).     
The long-term (90-year cycle) variation seen in the cycle-to-cycle 
variation (Fig.~\ref{fig4}) can also be seen in Fig.~\ref{fig6}(a).   
 We have used  the 
 3-MTIs for the sake of better statistics, but the aforementioned
patterns is also seen in the yearly data (figure is not shown here), in spite 
of the large uncertainties in the yearly values.

{\bf We would like to point out that the statistical significances of all the 
regression relations are tested by Student's `t' tests also. 
In all the cases the values of `t' are found to be high, 
 i.e., the values of their corresponding  
 probabilities  are found to be close to one
 (for example,  the interval 2008\,--\,2010
  has the smallest  number of data points  ${\rm N} = 163$. In  this case  
the value of 't' is found to be 7.375, whereas the cutoff value for the 
five percent significance level is only 1.654 and for one percent level
 it is 2.35).}  

Fig.~\ref{fig7} shows the plots of the  mean slope values  in 3-MTIs--i.e., 
the data shown in Fig.~\ref{fig6}(a)--versus the  year of the 
solar cycles, 11\,--\,24 (data are available only for four years of Cycle~11 and only for 3 years of Cycle~24). 
As can be seen in this figure, the pattern of the mean variation of the slope
(the closed circle-solid curve) suggests a slight increasing trend
   during the 
rising phases and a slight decreasing trend during  the decay phases 
of a majority of the solar cycles. However, 
the overall spread in the data points is 
very large, particularly in the beginnings and the endings of 
the cycles. That is,  the variations during the different solar cycles 
highly differ from the mean pattern, 
indicating the $\approx$ 11-year 
periodicity is very weak/absent in the slope. 

Overall we find  that  the  relationship  
$|D_{\rm n-1, n}|$  and  $A_{\rm n-1}$  is
reasonably consistent and reliable even in the cases of  
the  relatively small samples. 
It may be worth to note here that the average size of 
the spot groups considerably  varies  during a cycle. It also differs from 
cycle-to-cycle. Therefore, the temporal variation in the slope of the linear 
relationship may be mainly due to the dependence of the 
decay rate on the size/lifetime of the spot groups. 

We repeated all the above calculations for growth rate. 
The correlation  between the  growth rate, $G_{\rm n-1, n}$,   and 
the corresponding $A_{\rm n-1}$ is found to be low, $r= 0.39$, for the
 whole period data (with 60087 points). But  it is 
still found to be statistically significant from the above used all 
the significance tests 
(it should be noted that the values of n is different for
the growth and decay rates).
However, the correlation determined from the data of an individual cycle 
 is found to be still small and  statistically insignificant. Thus,  
the relationship between 
$G_{\rm n-1, n}$  and  $A_{\rm n-1}$ is considerably  inconsistent.
 Hence, it is not shown  here. 
There is also a considerable spread in Figs. 3 and 4. Therefore,  even the 
relationship between $|D_{\rm n-1, n|}$  and  $A_{\rm n-1}$ found above, 
is only suggestive rather than compelling. 

\section{Conclusions and discussion}
We analysed a large and reliable sunspot group data and found that
the total amounts of  growth
and  decay of spot groups whose life times $\ge 2$ days in a given
 time interval (say one-year)
 well correlate to the amount of activity in the same interval.
 We have also found that there exist a reasonably good correlation and
 an approximate linear 
relationship between  the  logarithmic values of the decay rate 
 and 
 area of the spot group at the first day of the corresponding 
consecutive days, largely suggesting that a large/small area (magnetic flux) 
decreases in a faster/slower rate.  
There exists a long-term (about 90-year) variation  in  the slope of this  
 relationship. 

 The growth of spot groups (enhancement of magnetic flux) may  have a
large contribution from the emergence of new magnetic flux.
 Hence, the growth of spot groups may be mostly   related to the generation
mechanism of solar activity, whereas for the decay process
magnetic reconnection may be the main mechanism.
 Magnetic reconnection is believed to be the main mechanism 
behind the solar energetic phenomena such as the solar flares, coronal 
mass ejection, etc..  Thus, the solar cycle variation in the decay of spot
 groups may have significant roles in the corresponding variations in the 
solar energetic phenomena, particularly solar flares which are mainly originate at sunspot locations.

{\bf The relationship (\ref{eqn4}) (see foot note of the   
$S_G - S_A$ and $S_D - S_A$ relations in Sec.~3.1)
largely implies  a   change
 in the topology of surface magnetic flux   
in every 10 days, which 
 is consistent with the following 
results/inferences:
\citet{Hlb81} analysed Mt. Wilson magnetograph data during the period
 1967 to mid-1980 and found that the rate at which the
magnetic flux appears on the
 Sun is sufficient to create all the flux that is seen at the solar surface
 within a period of about 10 days. The magnetic structures of the sunspot
groups may rise from near the bottom of the convention zone to
the surface in about 10 days~\citep{jg97}. In addition, a 
$\approx$~10-day periodicity seems to
be predominantly present in  the solar surface differential rotation, total solar irradiance (TSI), 
 solar coronal holes, solar wind, auroral electron power,
 and geomagnetic parameters~\citep[see][and references therein]{jj11b}, 
all these  may be 
related to the topology of the Sun's surface magnetic field.}  

It is known that
TSI  varies by about 0.1\% over the
 solar cycle.
A number of statistical
 models of TSI variability have been constructed on the basis of
inhomogeneities of surface magnetic field. These models helped to identify
the surface magnetic structures  are responsible for
 the variation
of  TSI and also provided widely accepted theoretical explanations.
According to these models, the dominant part of the TSI variations on 
time scales longer than about 
a day is caused by the evolution of solar surface magnetic features. 
 However, the exact mechanism behind the  TSI variation
  is not known~\citep{fou92,kuh99,sol05,dom09}.

TSI is well correlating with
the solar activity (for example with sunspot number)  and
 during the current unusually low and prolonged minimum  TSI also
 seems to be  unusually low~\citep{fro09}.
 Sunspots and  faculae act in opposite to modulate  TSI
 \citep{lea88,kri11}. That is, sunspots depress the solar 
energy output,
whereas faculae enhance it. However, the exact reason for the
correlation between solar activity and TSI is not yet known.

The growth and decay of sunspot
groups play key roles in the TSI  variations~\citep{wil81}.
The growth of spot groups may  enhance the  blocking of the
energy output. The decay  may  enhance the
energy output and
  mostly responsible for
   TSI  to
correlate with sunspot activity, $i.e.$, for the coherent relationship between
the activity and TSI variations.
Besides the convection,   the
 merdional flows
 may play significant roles in the magnetic reconnection, hence
in the  decay of spot groups (in general all kinds of active regions).

 \acknowledgments
The author thanks to one  anonymous referee for  critical review 
and to the other one 
for useful comments and suggestions.

\end{document}